# Symmetry and Numerical Solutions for Systems of Non-linear Reaction Diffusion Equations

Sanjeev Kumar* and Ravendra Singh
Department of Mathematics,
(Dr. B. R. Ambedkar University, Agra),
I. B. S. Khandari, Agra-282002
Email Id: *sanjeevibs@yahoo.co.in and ravendraibs@gmail.com

**Abstract:**
Many important applications are available for nonlinear reaction-diffusion equation especially in the area of biology and engineering. Therefore a mathematical model for Lie symmetry reduction of system of nonlinear reaction-diffusion equation with respect to one-dimensional Algebra is carried out in this work. Some classes of analytical and numerical solutions are obtained and expressed using suitable graphs.

## 1. Introduction:

Transports of molecular oxygen from the blood plasma to the living tissue of the skeletal muscle or brain across the capillary walls are nowadays very important topic. Several questions arise in our mind like (i) what factors affect the supply of oxygen tissue cell respirations? (ii) what happens when we inhale oxygen at low concentration (iii) what is the influence of axial and redial diffusion of oxygen in blood, oxygen diffusivity in tissue etc. Whenever we talk about the problem of diffusion, parallel we have to discuss about the diffusion or diffusion reaction equation. Diffusion reaction equations are then bifurcated into two ways, linear and nonlinear diffusion reaction equations. This problem is to obtain some different solutions with respect to a coupled nonlinear reaction diffusion equations.

Coupled systems of nonlinear diffusion equations have many important applications in mathematical physics, chemistry and biology. In population biology, the reaction terms models growth and the diffusion term accounts for migration. The classical diffusion term originates from a model in physics. Recent research indicates that the classical diffusion equation is inadequate to model many real situations, where a particle plume spreads faster than the classical model predates and may exhibit significant asymmetry. This situation is called anomalous diffusion.

Now as far as the solution of such equations is concern, we are having several approaches. Cherniha et al [6] worked on Lie symmetry of nonlinear multidimensional diffusion equation. Although symmetry is very much important therefore Nikitin et al [8] worked on the system of reaction diffusion equations and their symmetry properties. Tetyana [10] discussed about both symmetry and solution for system of nonlinear reaction diffusion equations. Lot of applications are available for such equations, Sharma et al [5] worked on the effect of nonzero bulk flow and non-mixing on diffusion with variable transfer of solute while Central et al

[11] discussed about the spatial ecology via reaction-diffusion equations, while Kumar et al [12] worked on a computational study of oxygen transport in the body of living organism.

## 2. Mathematical Model:

In the present model we consider a general system of nonlinear diffusion equation of the following form –

$$\frac{\partial U}{\partial t} - \frac{\partial^2}{\partial x^2}(a_{11}U + a_{12}V) - F = 0 \quad \ldots (1a)$$

$$\frac{\partial V}{\partial t} - \frac{\partial^2}{\partial x^2}(a_{21}U + a_{22}V) - G = 0 \quad \ldots (1b)$$

where $a_{11}, a_{12}, a_{21}, a_{22}$, are constant parameter with relation $a_{11} a_{12} - a_{21} a_{22} \neq 0$, and $F, G$ are the function of $U$ and $V$, while $U$, V are the functions dependent on $t$ and $x$.

An investigation of the equations (1) can be undertaken within the framework of the classical Lie algorithm which reduces the problem to determine the exact and numerical solution. The classical Lie symmetries of partial differential equations and classical symmetries of systems of two nonlinear diffusion equations with n+1 independent variable $t, x_1, \ldots, x_n$ were described. All possible non-linearties $F, G$ and the corresponding group generators were found. In the present work, using the results obtained in [7], we carried out symmetry reduction of equation (1) with respect to one-dimensional symmetry algebras and $F, G$ are defined up to arbitrary functions.

## 3. Symmetry Reductions:

If $\varphi_1$ and $\varphi_2$ are arbitrary function of $U$ and $a_{11}=a_{22}=a$, $a_{12}=0$, $a_{21}=b$, then a system of type (1) may be considered as:

$$\frac{\partial U}{\partial t} = a\frac{\partial^2 U}{\partial x^2} + \exp(\lambda_1 (U-V))\varphi_1 U \quad \ldots (2a)$$

$$\frac{\partial V}{\partial t} = b\frac{\partial^2 V}{\partial x^2} + \exp(\lambda_2 (U-V))(\varphi_1 V + \varphi_2) \quad \ldots (2b)$$

Now if the Greek letters $\alpha, \beta \in R$ denote the arbitrary coefficient, and $D_i$, $X_0$, $\hat{B}$ are various type of dilatation and special transformation generator, then some of the operators consider for this model, are as follows:

$$\left. \begin{array}{l} X_0 = \alpha\frac{\partial}{\partial t} + \beta\frac{\partial}{\partial x}, \quad D_1 = 2t\frac{\partial}{\partial t} + x\frac{\partial}{\partial x} - \frac{2}{\lambda}\hat{B}, \quad \hat{B} = b_{ij}u_b\frac{\partial}{\partial u_a} \\ \\ D_2 = 2t\frac{\partial}{\partial t} + x\frac{\partial}{\partial x} - \frac{2}{\lambda}\left(\frac{\partial}{\partial U} - 2nU\frac{\partial}{\partial V}\right), \quad D_3 = 2t\frac{\partial}{\partial t} + x\frac{\partial}{\partial x} - \frac{2}{\lambda}P_\alpha\frac{\partial}{\partial u_\alpha} \end{array} \right\} \quad \ldots (3)$$

where $b_{ij}$ are the elements of the 2*2 matrix B, and $\lambda, n$ are parameters used in the definitions of nonlinear term, and B, which have been specified by

$$\mathbf{B} = \begin{pmatrix} 0 & 0 \\ 1 & 0 \end{pmatrix} \quad \ldots (4)$$

The system of (2) permits the following symmetry operator:

$$X = X_0 + \upsilon D_1, \quad \ldots (5)$$

Now using Lie algorithm, the corresponding solution is:

$$U = w_1(z) \quad \ldots (6a)$$

$$V = -\frac{2}{\lambda} \ln(vx + \beta) + w_2(z) \quad \ldots (6b)$$

$$z = \frac{2(vx + \beta)^2}{2vt + \alpha}. \quad \ldots (6c)$$

Now if we substitute the solution (3) into (2), then we come to the following reduced equations, which is a system of ordinary differential equation.

$$2vz^2\dot{\omega}_1 + 4v^2 z\dot{\omega}_1 + 8v^2 az^2 \ddot{\omega}_1 = -2\exp(\lambda(\omega_1 - \omega_2))\varphi_1\omega_1 \quad \ldots (7a)$$

$$vz^2\dot{\omega}_2 + \frac{v^2 b}{\lambda} + 2v^2 bz\dot{\omega}_2 + 4v^2 bz^2 \ddot{\omega}_2 = -\exp(\lambda(\omega_1 - \omega_2))(\varphi_1\omega_2 + \varphi_2) \quad \ldots (7b)$$

**4. Numerical Solutions:**

Consider the following system (2) for the numerical solution

$$\frac{\partial U}{\partial t} = a\frac{\partial^2 U}{\partial x^2} + \exp(\lambda_1(U - V))\varphi_1 U$$

$$\frac{\partial V}{\partial t} = b\frac{\partial^2 V}{\partial x^2} + \exp(\lambda_2(U - V))(\varphi_1 V + \varphi_2)$$

We solve the above partial differential equation through the MATLAB 6.0, and consider following value of the represent the parameters:

$$\begin{array}{ll} a = 0, & \lambda_1 = 5.73, \quad \varphi_1 = 0.5 \\ \text{and } b = 1, & \lambda_2 = -11.47 \quad \varphi_2 = 1.5 \end{array} \quad \ldots (8)$$

We have taken x=0 to 1 and t=0 to 2 sec. Figure (1) and (2) represents the value of U and V respectively with respect to time t and distance x.

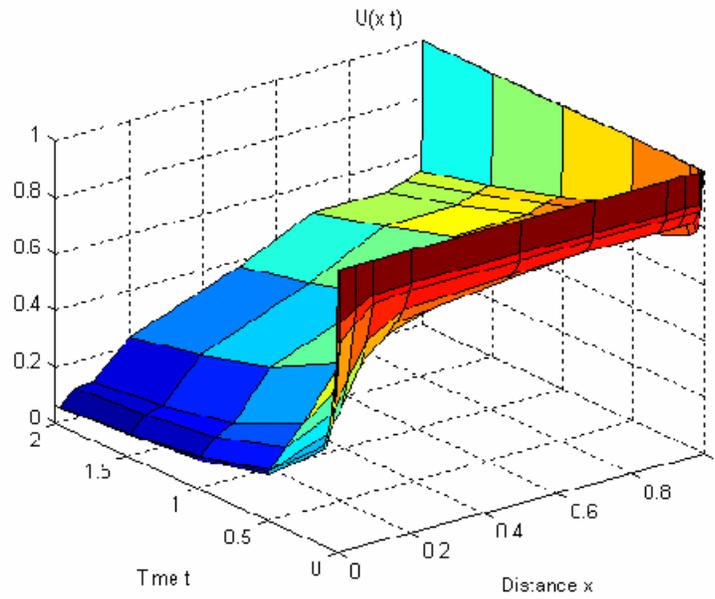

***Figure 1*** shows the function U at different value of distances *x* and time *t*. The value of *U* is decreasing with respect to time t and distance x.

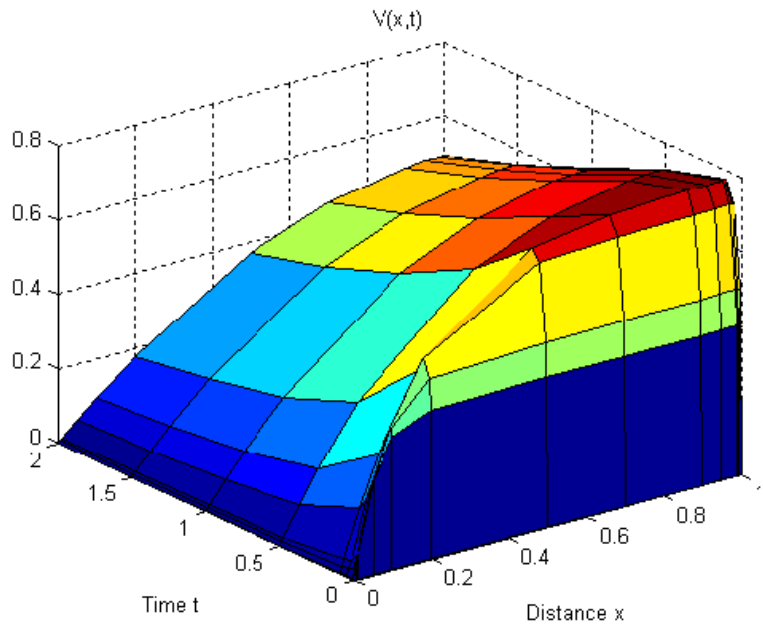

***Figure 2*** shows the function V at different value of distances *x* and time *t*. The value of *U* is decreasing with respect to time t and distance x.

## 5. Conditional symmetry and exact solution:

Thus we presented reduction of equation (1) using their classical symmetry found in [7]. In this section we present exact solution of equation (1) found by conditional symmetry reduction.

$$\frac{\partial U}{\partial t} - \frac{\partial^2 U}{\partial x^2} = U^3 \varphi_1 \qquad \ldots (9a)$$

$$\frac{\partial V}{\partial t} - \frac{\partial^2 V}{\partial x^2} = V^3 \varphi_2 \qquad \ldots (9b)$$

where $\varphi_1$ and $\varphi_2$ are function of $\frac{U}{V}$.

Conditional symmetry operator:

$$X = \frac{\partial}{\partial t} - \frac{3}{x+k_1}\frac{\partial}{\partial x} - \frac{3}{(x+k_1)^2}\left(U\frac{\partial}{\partial U} + V\frac{\partial}{\partial V}\right). \qquad \ldots (10)$$

The ansatz

$$U = (x+k_1)\omega_1(z) \qquad \ldots (11a)$$
$$V = (x+k_1)\omega_2(z) \qquad \ldots (11b)$$
$$z = \frac{1}{2}x^2 + k_1 x + 3t \qquad \ldots (11c)$$

Reduce equation (4) to the following system:

$$\ddot{\omega}_1 + \varphi_1 \omega_1^3 = 0 \qquad \ldots (12b)$$

$$\ddot{\omega}_2 + \varphi_2 \omega_2^3 = 0 \qquad \ldots (12b)$$

where $\varphi_1$ and $\varphi_2$ are function of $\frac{\omega_2}{\omega_1}$.

Depending on the form of the functions $\varphi_1$, $\varphi_2$, we receive different solution the system.

1) $\varphi_1 = a > 0$, $\varphi_2 = b < 0$, where $a$ and $b$ are constants.

$$U(x,t) = \frac{\sqrt{2a}}{2a}(x+k_1)sd\left(\frac{1}{2}x^2 + k_1 x + 3t; \frac{1}{2}\sqrt{2}\right) \qquad \ldots (13b)$$

$$V(x,t) = -\frac{\sqrt{-2b}}{b}(x+k_1)ds\left(\frac{1}{2}x^2 + k_1 x + 3t; \frac{1}{2}\sqrt{2}\right) \qquad \ldots (13b)$$

2) $\varphi_1 = a > 0$, $\varphi_2 = 0$

$$U(x,t) = \frac{\sqrt{2a}}{2a}(x+k_1)sd\left(\frac{1}{2}x^2 + k_1 x + 3t; \frac{1}{2}\sqrt{2}\right) \qquad \ldots (14b)$$

$$V(x,t) = (x+k_1)\left[\left(\frac{1}{2}x^2 + k_1 x + 3t;\right)C_1 + C_2\right]. \qquad \ldots (14b)$$

## 6. Conclusions:

Through section 4, we have the numerical results for the U and V at different values of distances *x* and time *t*. The function *U* is decreasing with respect to time t and distance x and the function *V* is increasing with respect to time t and distance x. In work paper we have presented a post processing algorithm that the system of nonlinear reaction diffusion equation solvable in exact and numerical form.

Such type of linear and nonlinear reaction, diffusion, and reaction-diffusion equations may be used to model spatio-temporal processes in physical systems, like astrophysical fusion plasmas. We suggest the use of such models or systems to describe the natural structures as patterns in chemistry, ecology, biology and physics or the signal behavior in the neural context